\documentclass[twocolumn]{aastex62}
\usepackage{amsmath}
\usepackage{CJK}

\newcommand{\rp}{{R_{\rm p}}}
\newcommand{\snr}{{\rm S/N}}
\newcommand{\cdpp}{{\rm CDPP}_{\rm 6hr}}
\defcitealias{Weiss:2018}{W18}

\shorttitle{Patterns in \emph{Kepler} multi-planet systems}
\shortauthors{Wei Zhu}

\begin{document}
\begin{CJK*}{UTF8}{gbsn}

\title{On the patterns observed in \emph{Kepler} multi-planet systems}

\author{Wei~Zhu (祝伟)}
\affil{Canadian Institute for Theoretical Astrophysics, University of Toronto, 60 St. George Street, Toronto, ON M5S 3H8, Canada}
\correspondingauthor{Wei Zhu}
\email{weizhu@cita.utoronto.ca}

\begin{abstract}
Recent studies claimed that planets around the same star have similar sizes and masses and regular spacings, and that planet pairs usually show ordered sizes such that the outer planet is usually the larger one. Here I show that these patterns can be largely explained by detection biases. The \emph{Kepler} planet detections are set by the transit signal-to-noise ratio ($\snr$). For different stellar properties and orbital period values, the same $\snr$ corresponds to different planetary sizes. This variation in the detection threshold naturally leads to apparent correlations in planet sizes and the observed size ordering. The apparently correlated spacings, measured in period ratios, between adjacent planet pairs in systems with at least three detected planets are partially due to the arbitrary upper limit that the earlier study imposed on the period ratio, and partially due to the varying stability threshold for different planets. After these detection biases are taken into account, we do not find strong evidence for the so-called ``intra-system uniformity'' or the size ordering effect. Instead, the physical properties of \emph{Kepler} planets are largely independent of the properties of their siblings and the parent star. It is likely that the dynamical evolution has erased the memory of \emph{Kepler} planets about their initial formation conditions. In other words, it will be difficult to infer the initial conditions from the observed properties and the architecture of \emph{Kepler} planets.
\end{abstract}

\keywords{methods: statistical --- planetary systems --- planets and satellites: general}

\section{Introduction} \label{sec:introduction}

Almost all planets reside in multi-planet systems, and the physical and orbital properties of planets in the same system convey important clues about their formation and evolution.
To date the majority of the multi-planet systems were found by \emph{Kepler} \citep{Borucki:2010}.
\footnote{See a list of all known multi-planet systems at \href{http://exoplanetarchive.ipac.caltech.edu}{NASA Exoplanet Archive} \citep{Akeson:2013}.}
However, \emph{Kepler} could only detect those planets that transited their hosts and had transit signals above some certain noise level. This limits our knowledge about the detected multi-planet systems and complicates the theoretical interpretation.

As \emph{Kepler} observations provide directly the transit planet-to-star radius ratio, the relative sizes of planets inside the same system can be easily examined. Based on data from the first four months of observations, \citet{Lissauer:2011} pointed out that adjacent planets were likely to have very similar radii, as most of the pairs show $R_{\rm p,in}/R_{\rm p,out}\approx 1$. Here $R_{\rm p,in}$ and $R_{\rm p,out}$ are the radii of the inner and the outer transiting planets (tranets), respectively. This feature was further studied in \citet{Ciardi:2013}, and the authors reported that most ($>60\%$) of the multi-planet systems found by \emph{Kepler} appeared to have this size-location correlation: the outer planet was larger than the inner one.
\footnote{\citet{Ciardi:2013} argued that this result only applied to planet pairs in which at least one was approximately Neptune-sized or larger. However, without this constraint the results are qualitatively similar in the statistical sense. See their Figures 4 and 11.}
In both studies, the authors compared the observed and the simulated radius ratio distributions to determine the statistical significance of their finding. Their simulated distributions were produced by randomly drawing radii from the \textit{observed} radius distribution and going through customized signal-to-ratio (S/N) cuts. As I will explain later, this approach does not capture all the detection biases.

Later follow-up efforts that lead to better characterizations of \emph{Kepler} stars allow comparisons of planetary parameters across systems. Recently, \citet[][hereafter W18]{Weiss:2018} claims that planetary systems are like ``peas in a pod,'' namely that the planets orbiting around the same host have similar sizes and regular spacings. They took the large sample of \emph{Kepler} multi-planet systems whose parameters were refined by the California-\emph{Kepler} Survey (CKS, \citealt{Petigura:2017,Johnson:2017}), sorted the \emph{CKS planets}
\footnote{A ``CKS planet'' is a planet that transits the host, is detectable by \emph{Kepler}, and is included in the CKS sample. CKS planets are almost certainly valid planets, but because of the geometric transit probability and \emph{Kepler} detection sensitivity, the CKS planets are not necessarily all the planets in those systems.}
in the same systems according to their orbital periods, and computed the correlation between sizes of neighboring CKS planets. They then quantified the significance of this correlation through bootstrap tests and found that the observed correlation could not be explained by randomly resampling the observed size distribution. The procedure was similar for the spacings between planets.

A later
\footnote{The \citetalias{Weiss:2018} work was posted on the arXiv pre-print server before the \citet{Millholland:2017} work.}
work by \citet{Millholland:2017} adopted a similar statistical approach and further claimed that the masses of planets inside the same system, given by \citet{Hadden:2017} from analyzing the transit timing variations (TTVs, \citealt{Agol:2005,Holman:2005}), should also be similar.
This, together with the aforementioned trends about radius and spacing, was summarized as the ``intra-system uniformity" \citep{Millholland:2017}.

However, an issue that was overlooked in these studies \citep{Lissauer:2011,Ciardi:2013,Weiss:2018,Millholland:2017} is the detection threshold. Below I use the transit detection as an example, but the idea applies to TTV mass measurements as well (see Section~\ref{sec:discussion}). The \emph{Kepler} transit search pipeline requires a nominal minimum S/N of $7.1$; for the planet sample used in \citetalias{Weiss:2018}, a higher (S/N $=10$) threshold was used. The detection completeness depends strongly on the S/N at low end \citep[e.g.,][]{Fressin:2013,Thompson:2018}. Given the dominating contribution from smaller planets and weaker transits \citep[e.g.,][]{Hsu:2019}, the S/N values of \emph{Kepler} transit detections as a result appear to pile up toward the detection threshold (e.g., Figure~1 of \citealt{Ciardi:2013}) and are not affected by the variations of stellar parameters or noise levels. For a certain S/N threshold, the resulting planet radius threshold depends on the orbital period as well as the host properties. This variation was not fully taken into account by those referenced works in generating simulated parameters.

A more proper way is fully forward modeling the detection and selection processes from the \emph{intrinsic} planetary (radius or mass) distribution. Instead of randomly drawing parameters from the observed distribution, one should draw from the intrinsic distribution and then apply the same detection criteria (e.g., S/N cut) on these simulated planets. This process requires knowing the intrinsic planet distribution function and having access to the automated \emph{Kepler} detection pipeline. It is further complicated by the fact that the planet distribution function is period dependent \citep[e.g.,][]{Dong:2013,Hsu:2019} and possibly multiplicity dependent, and that the \emph{Kepler} detection efficiency is weakly multiplicity dependent \citep{Zink:2019}.

\begin{figure}
\epsscale{1.1}
\plotone{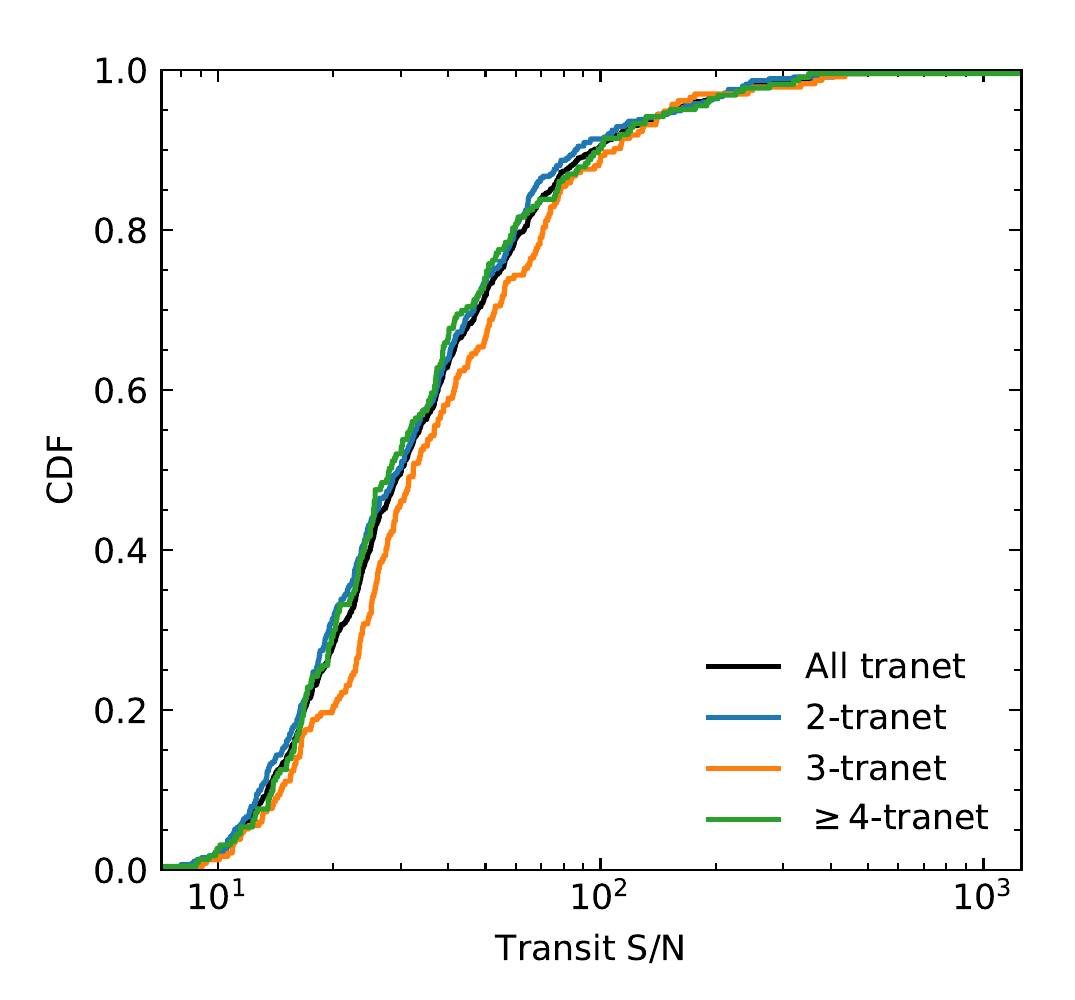}
\caption{Cumulative distributions of transit S/N. Different transit multiplicities are shown in different colors, and the overall sample is shown in black. There is no clear dependence of S/N on the transit multiplicity.
\label{fig:snr_cdf}}
\end{figure}

\begin{figure*}
\epsscale{1.}
\plotone{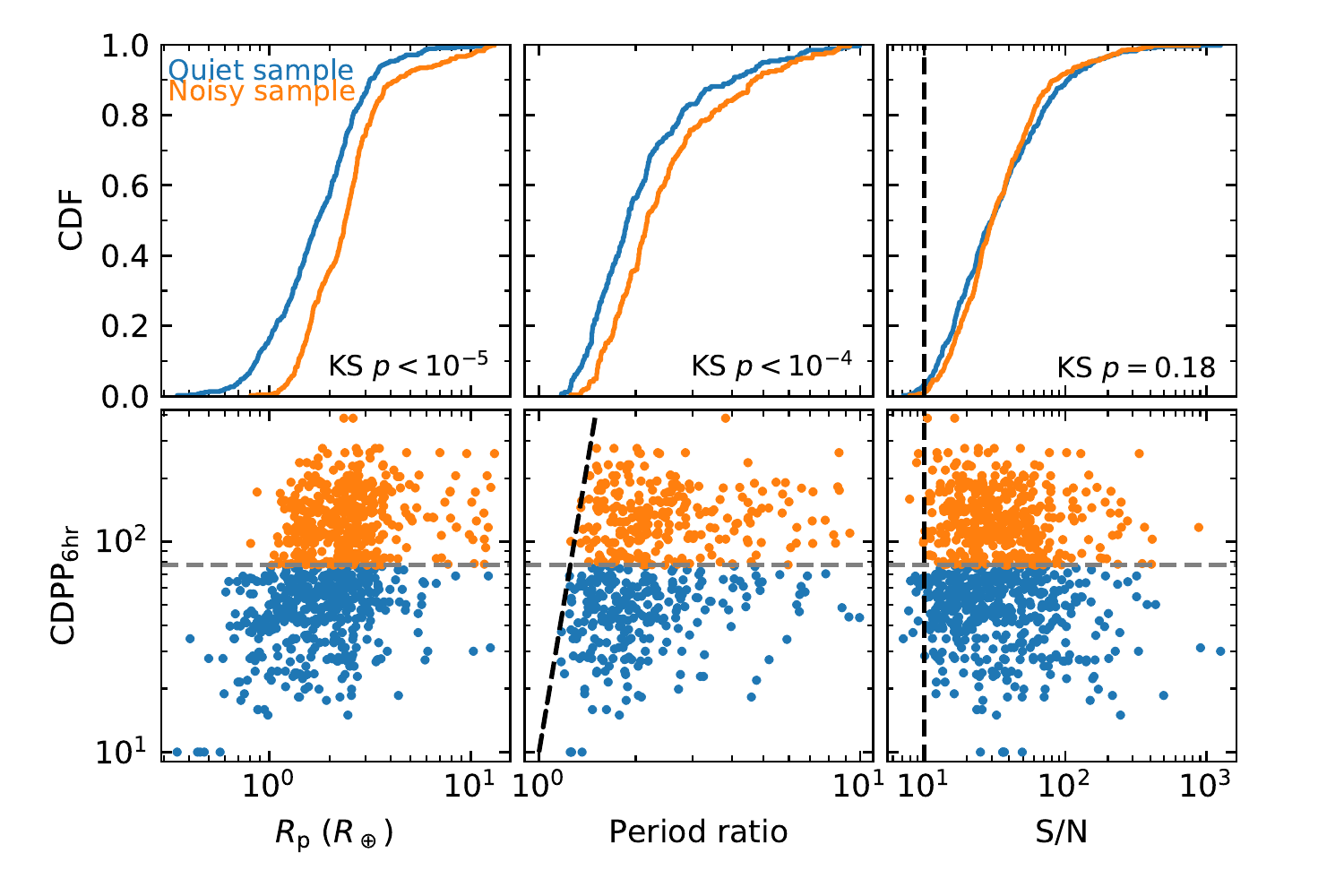}
\caption{Lower panels show the noisy level of the planet host, characterized by $\cdpp$, versus planet radius $R_{\rm p}$ (lower left), planetary period ratio (lower middle), and signal-to-noise ratio (S/N) of the transit detection (lower right). The gray dashed horizontal lines mark the median $\cdpp$, based on which the sample is divided into two. In the upper panels we compare the cumulative distribution functions (CDF) of the individual parameter ($R_{\rm p}$, period ratio, and S/N) from the two samples, and the two-sample KS test $p$ values are indicated. The radius and period ratio distributions of the two samples are statistically different. In the case of period ratio (lower middle panel), we also mark with the black dashed line the smallest $\cdpp$ for various period ratios. The S/N distributions are very similar between the two samples. Nearly 50\% of planet detections have $\snr<30$, i.e., only a factor of three above the detection threshold ($\snr=10$, as marked by the vertical dashed line).
\label{fig:sample}}
\end{figure*}

There is a shortcut that circumvents these problems. In the full forward modeling approach, one generates synthetic planetary systems, adds stellar noises, passes them to the \emph{Kepler} detection pipeline, decides which planets are detectable, and finally performs statistical analyses on the simulated detections. Through this whole process planetary physical parameters are converted into transit observables and the detectability of individual planet is controlled by the transit S/N. Given the central role of transit S/N, we can directly start from this parameter as a shortcut to the full forward modeling approach. As shown in Figure~\ref{fig:snr_cdf} and the upper right panel of Figure~\ref{fig:sample}, the S/N distribution is independent of the transit multiplicity and the stellar properties. Therefore, we can randomly draw detections from this universal S/N distribution, derive planetary parameters, and perform the same statistical analysis as we do on the real data. As the relation between S/N and planetary radius indicates (Equation~\ref{eqn:snr}), when the minor contribution from orbital period is ignored, correlated sizes will definitely lead to correlated S/N values, but correlated S/N values does not necessarily mean correlated sizes because of the same stellar size and stellar noise level that two adjacent planets share. Therefore, by randomly sampling the S/N distribution and thus assuming no correlation in S/N values, I am being more generous than just assuming no size correlation.

In this work, I apply this more robust statistical approach to study the patterns observed in \emph{Kepler} multi-planet systems. I describe the planet sample in Section~\ref{sec:sample} and explain the basic idea in section~\ref{sec:idea}. Then in Sections~\ref{sec:radius} and \ref{sec:pratio} I discuss the issues involved in the ``peas in a pod'' claim. The size-location correlation is re-evaluated in Section~\ref{sec:size_hierarchy}. Finally, I briefly comment on the intra-system mass uniformity claim and then discuss the results in Section~\ref{sec:discussion}.

\section{Sample} \label{sec:sample}

\begin{figure*}
\epsscale{1.1}
\plotone{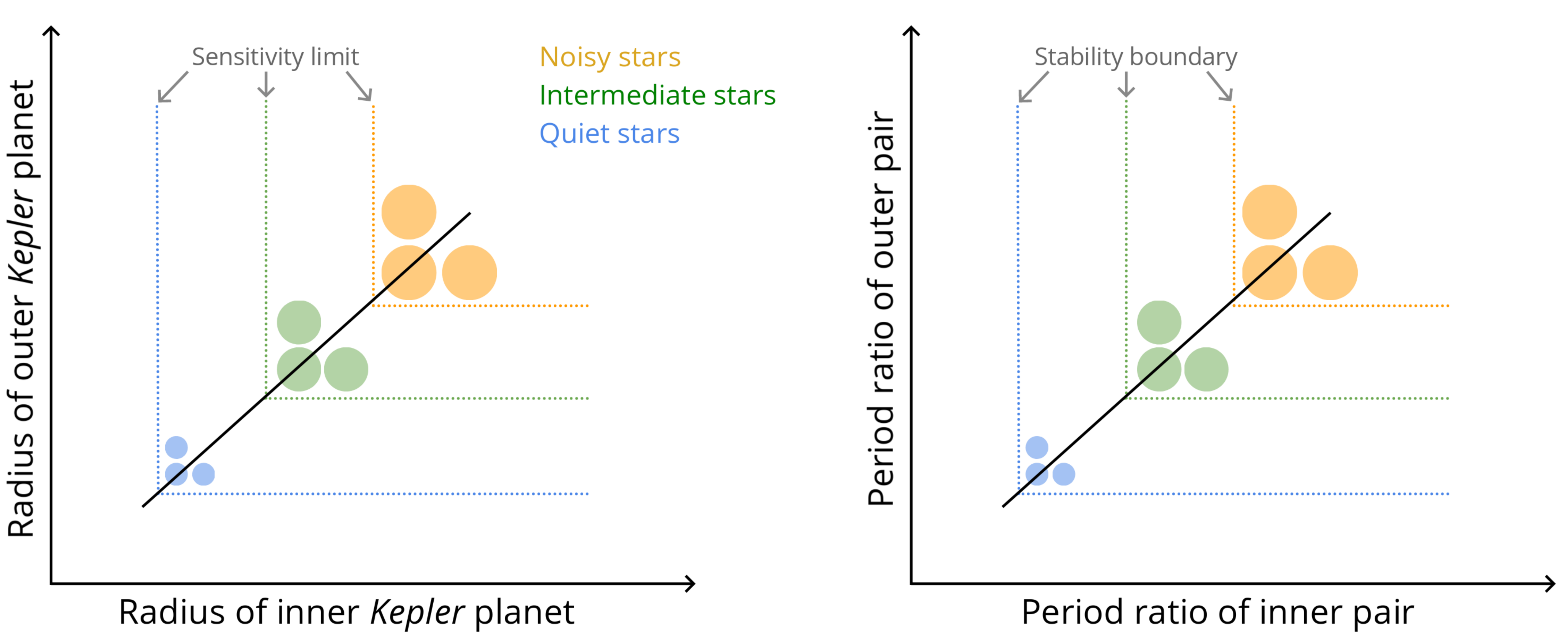}
\caption{A schematic view of how an inhomogeneous stellar sample will lead to the size (left panel) and spacing (right panel) correlations. For stars with a certain noise level, only planets above a certain size are detectable. This can be seen in the lower left panel of Figure~\ref{fig:sample}. Given that there are more smaller planets than larger ones, detected planet pairs tend to cluster around the corner that is defined by the vertical and horizontal sensitivity limits. For different stellar samples (as measured by their noise levels), these clusterings appear at different locations along the diagonal line, and thus a collection of these planets will naturally show a size correlation (left panel). Similarly, the stability boundary in the measure of period ratio decreases with decreasing planet sizes, as seen in the lower middle panel of Figure~\ref{fig:sample}. Because of the strong bias in orbital period of transit detections, observable planet triplets will tend to cluster around the corner of the stability boundary. For different stellar samples (as measured by the noise level), these clusterings appear at different locations along the diagonal line, and thus a collection of these planet triplets will naturally show a spacing correlation (right panel).
\label{fig:schematic}}
\end{figure*}

\begin{figure*}
\epsscale{1.2}
\plotone{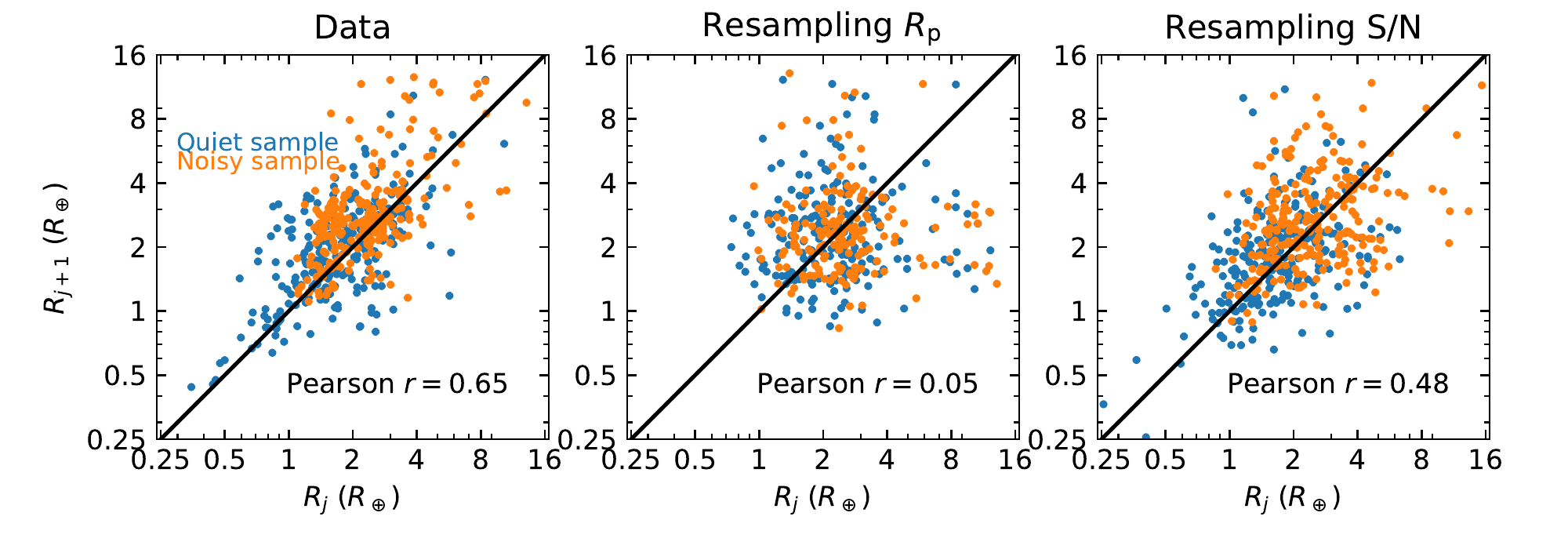}
\caption{Comparisons between radii of adjacent planets ($R_{\rm j}$ and $R_{\rm j+1}$). In the left panel is the distribution of planet pairs in the CKS multi-planet sample. In the middle and the right panels are the synthetic planet pairs generated in two different approaches: resampling $\rp$ and resampling S/N. The former was used in \citetalias{Weiss:2018}. Planet pairs from two stellar samples (quiet and noisy) are plotted with different colors to highlight the difference. The correlation coefficient between $\log{R_j}$ and $\log{R_{j+1}}$ is also indicated at the lower right corner of each panel.
\label{fig:radius}}
\end{figure*}

I use the same multi-planet sample as in \citetalias{Weiss:2018}. This sample includes 909 CKS planets in 355 multi-planet systems. The parameters of the individual planets and of their hosts were provided in Table~1 of \citetalias{Weiss:2018}. Of relevance to this study are the stellar mass $M_\star$, stellar radius $R_\star$, 6 hr Combined Differential Photometric Precision ($\cdpp$, a measure of the stellar noise level, \citealt{Christiansen:2012}), impact parameter $b$, planetary radius $\rp$, and orbital period $P$. This table did not include a column of S/N, but it can be easily computed by
\begin{equation} \label{eqn:snr}
{\rm S/N} = \frac{(\rp/R_\oplus)^2 \sqrt{{3.5~\rm yr}/P}}{\cdpp \sqrt{{6~\rm hr}/T}}\ .
\end{equation}
Here $T$ is the transit duration, given by
\begin{equation} \label{eqn:duration}
T = 13~{\rm hr} \left(\frac{P}{1~\rm yr}\right)^{1/3} \left(\frac{\rho_\star}{\rho_\odot}\right)^{-1/3} \sqrt{1-b^2}\ ,
\end{equation}
where $\rho_\star$ is the stellar mean density. Note that the Equation (2) of \citetalias{Weiss:2018} did not have the factor $\sqrt{1-b^2}$, which made their S/N overestimated, although this would only have a minor effect on the results.

I show in Figure~\ref{fig:sample} the stellar noise level $\cdpp$ versus three chosen parameters: planet radius $\rp$, period ratio of adjacent CKS planets, and the transit detection S/N. For demonstration purposes I divide the whole sample into two at the median $\cdpp$: the \textit{quiet sample} and the \textit{noisy sample}.

The S/N value dictates the significance of a transit detection and is a more fundamental observable in signal detections than planetary radius, which is not even a direct observable in transit light curve. Because S/N already takes into account variation of stellar noise level $\cdpp$ (Equation~(\ref{eqn:snr})), one does not expect the S/N distribution to be different between the quiet and the noisy samples. Indeed, a two-sample Kolmogorov-Smirnov (KS) test between the S/N distributions from two samples gives $p=0.18$, confirming that the S/N distribution is invariant to the variations of stellar properties. However, the radius distributions and the period ratio distributions from the two samples are different: the two-sample KS test gives $p<10^{-5}$ and $p<10^{-4}$ respectively. These differences are most prominent at small values of $\rp$ and period ratio. The difference in $\rp$ is due to the projection of the same S/N distribution into different stellar samples: smaller planets are more easily detected around more quiet stars. Then through the dynamical stability requirement, the difference in $\rp$ distributions propagates into the difference in period ratio distributions. See Section~\ref{sec:pratio} for more details.

Therefore, across the whole sample there is a universal S/N distribution but no universal radius distribution or period ratio distribution. The latter two were used in the bootstrap test by \citetalias{Weiss:2018}.

\section{The idea of this paper} \label{sec:idea}

Figure~\ref{fig:schematic} illustrates how the variation in detection threshold originating from a fixed S/N can lead to size and spacing correlations in observed planets.

First of all, one should know that there are more smaller planets than larger ones, at least down to \emph{Kepler}'s sensitivity limit, after the correction of detection bias \citep[e.g.,][]{Hsu:2019}. Therefore, if only planets above a certain size are detectable, then the detected planets will tend to pile up toward the detection threshold. In realistic missions such as \emph{Kepler}, the detection threshold is usually fixed in S/N because of its central role in signal detection. However, given the relation between planetary radius and transit S/N (Equation~(\ref{eqn:snr})) a fixed S/N threshold will lead to different radius thresholds for stars with different noise levels. Such a varying radius threshold will naturally lead to a correlation between the sizes of neighboring \emph{Kepler} planets. See the left panel of Figure~\ref{fig:schematic} for a simple illustration.

\citetalias{Weiss:2018} used bootstrap tests on planetary radii to study the significance of the size correlation. The underlying assumption behind their radius bootstrap test is that the radius distribution, $P(\rp)$, should be universal across different stellar subsamples. I have shown in the previous section that this assumption does not hold for the sample under investigation. Instead, the radius distribution depends on the stellar properties, specifically the noise level $\cdpp$. I denote this conditional radius distribution as $P(\rp|\cdpp)$ and note that
\begin{equation}
P(\rp) \neq P(\rp|\cdpp)\,.
\end{equation}
A proper radius bootstrap test should therefore be randomly drawing radii from the conditional radius distribution $P(\rp|\cdpp)$. Unfortunately, such a conditional radius distribution cannot be easily specified, but one can use the relation between $\rp$ and $\snr$ to further simplify the procedure. With other parameters the same (as is required by the bootstrap test), $\rp$ uniquely determines $\snr$. Thus randomly sampling $P(\rp|\cdpp)$ is equivalent as randomly sampling $\snr$ from $P(\snr|\cdpp)$ and then deriving $\rp$ from $\snr$. Recalling that the $\cdpp$ distribution is universal across different stellar subsamples,
\begin{equation}
P(\snr|\cdpp) = P(\snr)\,,
\end{equation}
one can further simplify the proper radius bootstrap test as randomly resampling the $\snr$ distribution $P(\snr)$. This is another justification of the $\snr$-resampling method that is used in this work.

Similar to the claimed size correlation, the claimed spacing correlation is also affected by the variation in detection threshold. With decreasing stellar noise, the minimum period ratio between adjacent planets also decreases, as shown in the lower middle panel of Figure~\ref{fig:sample}. This is probably because of the stability boundary, in the measure of period ratio, decreases with decreasing planet size and we refer to Section~\ref{sec:pratio} for a more detailed explanation. As the transit probability strongly biases toward small period values and thus small period ratios, the detectable planet pairs tend to pile up toward the smallest period ratio (i.e., the stability boundary). The variation of this stability boundary in an inhomogeneous stellar sample naturally leads to a spacing correlation. See the right panel of Figure~\ref{fig:schematic} for an illustration.

\section{On the radius uniformity} \label{sec:radius}

\begin{figure}
\epsscale{1.2}
\plotone{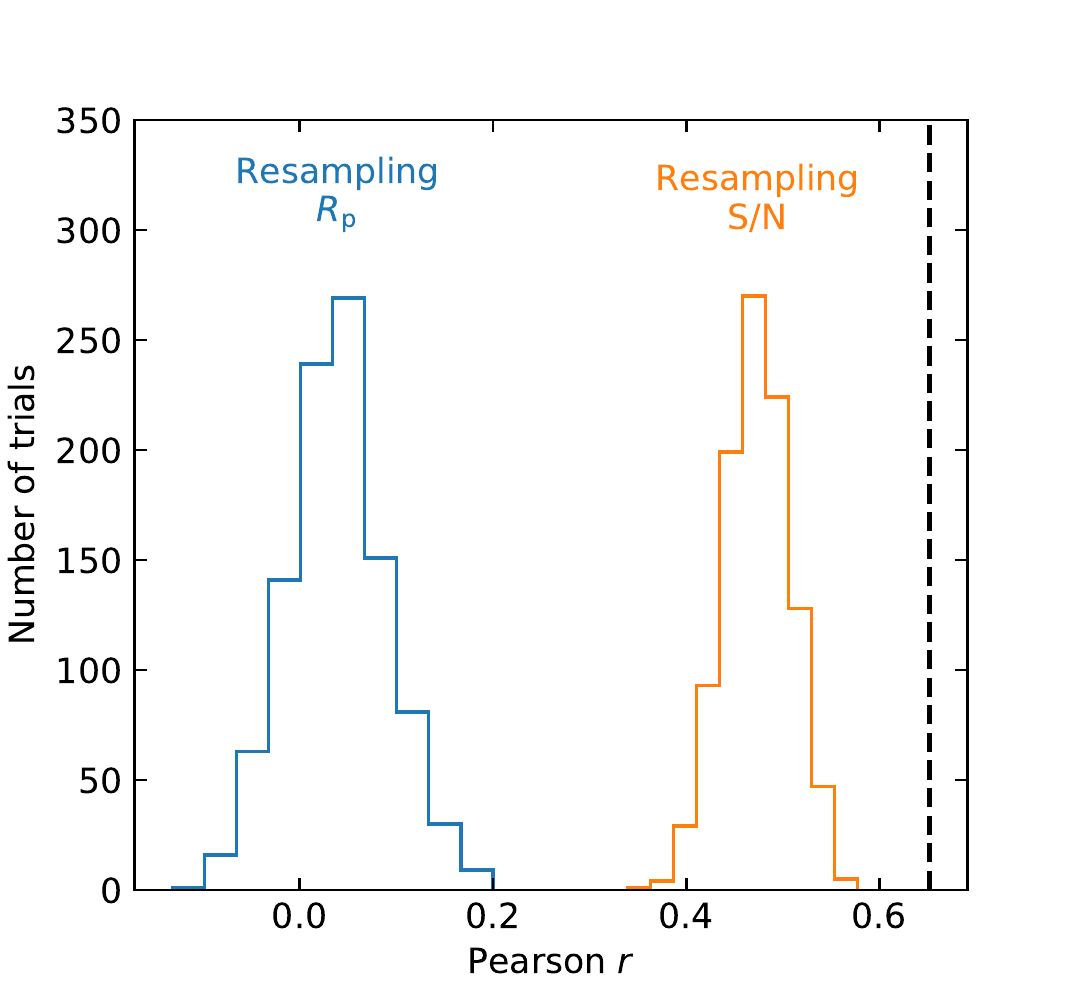}
\caption{The distributions of the Pearson $r$ coefficients from statistical tests. The blue and orange histograms are results from two different approaches, and the black dashed line indicates the measured correlation value ($r=0.65$). Note that resampling radius is a bootstrap method whereas resampling S/N is a forward modeling approach.
\label{fig:histograms}}
\end{figure}

\begin{figure*}
\epsscale{1.2}
\plotone{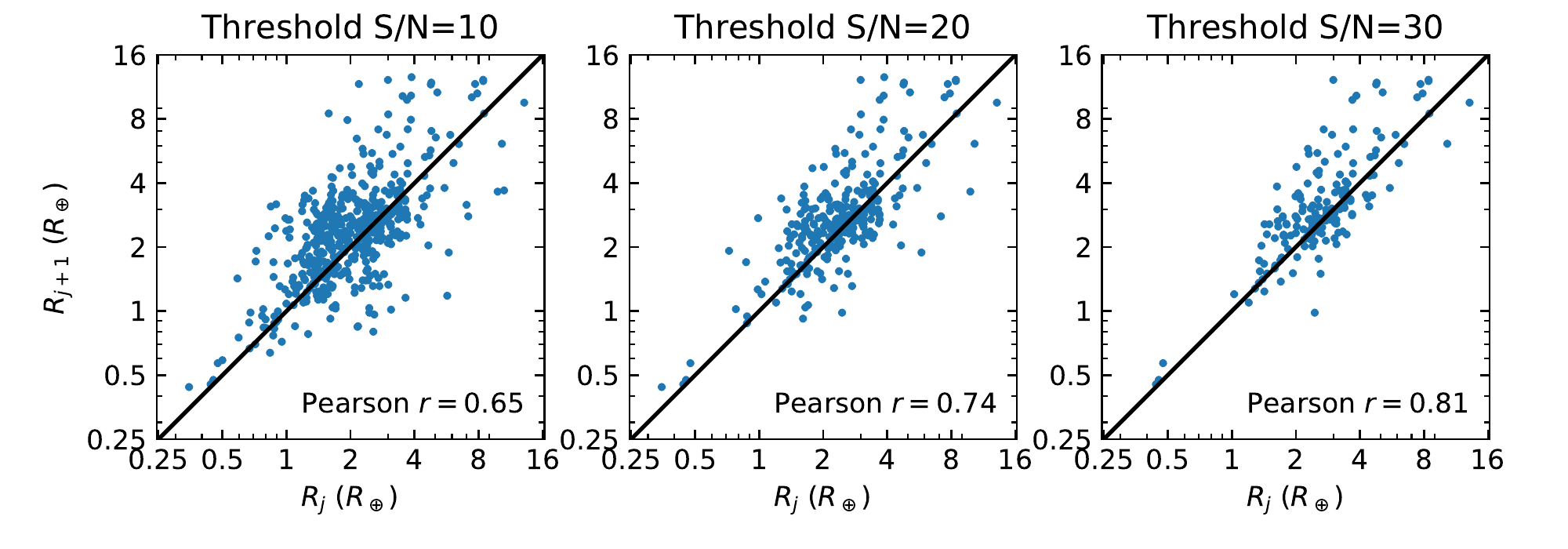}
\caption{The size correlation plots for different choices of threshold S/N values. Increasing the threshold S/N is equivalent to decreasing \emph{Kepler} sensitivity, which leads to increased size correlation. Extrapolating to lower S/N thresholds, this test suggests that a much better \emph{Kepler}-like mission which is sensitive to smaller planets will find a much weaker size correlation.
\label{fig:correlations}}
\end{figure*}

\begin{figure*}
\epsscale{1.1}
\plotone{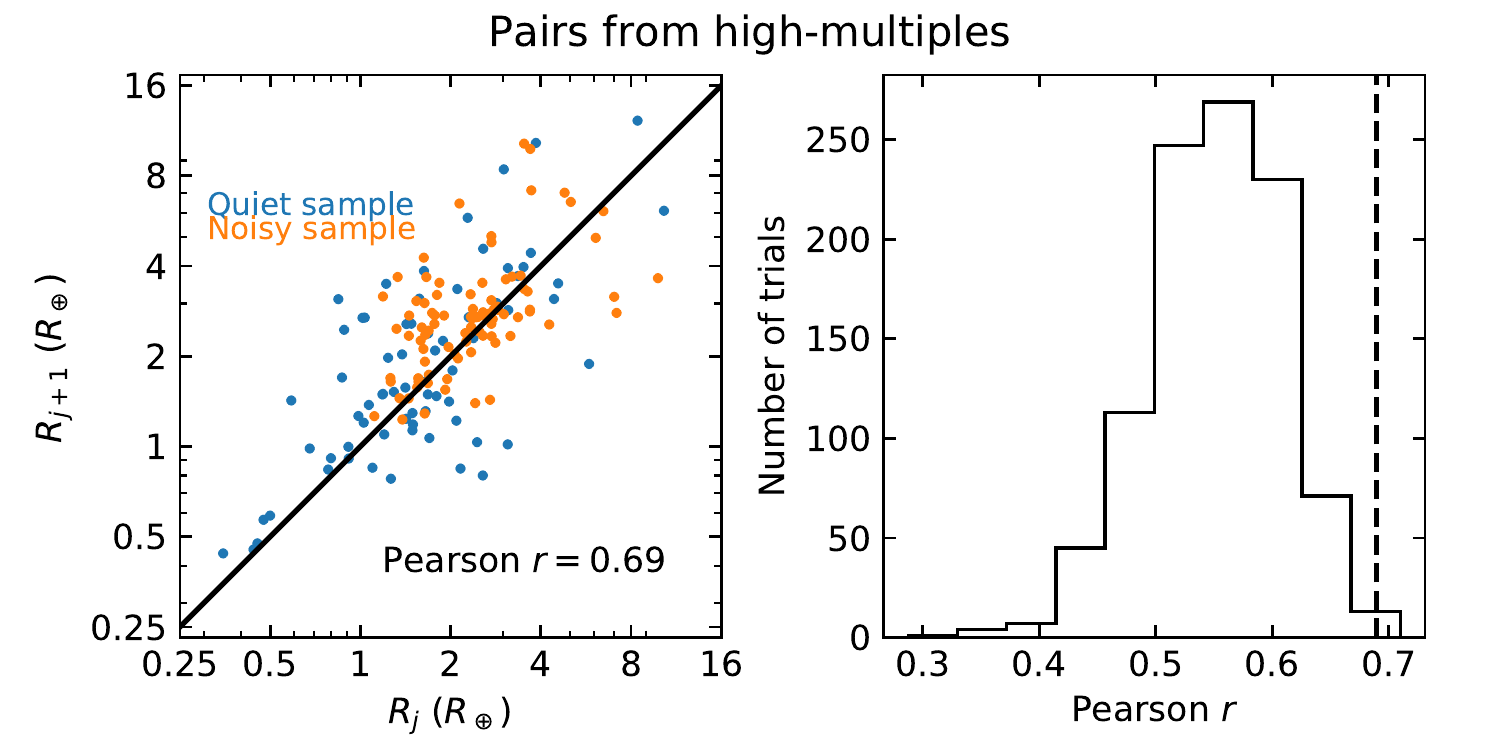}
\caption{The left panel shows the comparison between radii of adjacent planets ($R_{\rm j}$ and $R_{\rm j+1}$) from compact systems, defined as systems with at least 4 transiting planets. planet pairs from two stellar samples are marked with different colors. The correlation coefficient is given in the lower right corner. For these planet pairs, 1000 statistical tests is performed, in which the S/N (rather than $\rp$) is resampled, and the resulting Pearson $r$ values are shown as the histogram in the right panel. The observed correlation coefficient is not much different from the correlation we get in the statistical test.
\label{fig:compact}}
\end{figure*}

In the left panel of Figure~\ref{fig:radius} I show the radius of one CKS planet, $R_j$, versus the radius of the outer adjacent CKS planet, $R_{j+1}$. This is very similar to Figure 2 of \citetalias{Weiss:2018}. The correlation coefficient (Pearson $r$) between $\log{R_j}$ and $\log{R_{j+1}}$ of all planet pairs is $r=0.65$, consistent with the value reported in \citetalias{Weiss:2018}.

To assess the importance of this correlation \citetalias{Weiss:2018} generated synthetic planet systems in which the radius of each planet was randomly drawn from the overall radius distribution. An example realization following the procedure of \citetalias{Weiss:2018} is shown in the middle panel of Figure~\ref{fig:radius}. Note that in this plot there appears to be fewer sub-Earth-sized planets. This is because, following \citetalias{Weiss:2018}, simulated planets with S/N$<10$ are excluded. This step reduces the number of planets by nearly 20\%.

However, as discussed in Section~\ref{sec:sample} and illustrated in Figure~\ref{fig:sample}, the S/N distribution is more fundamental and universal than the radius distribution in transit signal detections. In particular, the radius distribution appears different for stars with different noise levels. This can also be seen in the left panel of Figure~\ref{fig:radius}, where I have differentiated the planet pairs from quiet and noisy samples. Note the similarity between this plot and the left panel of Figure~\ref{fig:schematic}. Following the reasoning in Section~\ref{sec:idea}, I therefore modify the bootstrap test of \citetalias{Weiss:2018}. Instead of resampling the $\rp$ distribution, I resample the S/N distribution and then, with other parameters unchanged, derive $\rp$ from Equation~(\ref{eqn:snr}).
\footnote{In practice, this can be easily achieved with the relation $R_{\rm p}^{\rm new} = \rp \sqrt{({\rm S/N})_{\rm new}/({\rm S/N})}$.}
Note that while I am bootstrapping transit S/N, I am essentially performing a forward modeling (see Section~\ref{sec:introduction} for the detailed explanation). The result from one random test is shown in the right panel of Figure~\ref{fig:radius}. In this simulated sample the planet pairs from two stellar samples show a systematic offset, a feature that is similar to the data (left panel). The radii of adjacent planets also show significant correlation, $r=0.48$.

I repeat the above statistical test for 1000 times, record all Pearson $r$ coefficients, and show their histogram in Figure~\ref{fig:histograms}. For comparison purposes I also produce the histogram of $r$ coefficients from 1000 bootstrap tests following the \citetalias{Weiss:2018} procedure (i.e., resampling $\rp$), and the resulting histogram peaks at $r\approx0$, similar to what \citetalias{Weiss:2018} had (see their Figure 5). By resampling on the more fundamental parameter $\snr$, I almost always reproduce, at least qualitatively, the observed size correlation, although with an average correlation coefficient $r\approx0.5$ this effect alone cannot explain quantitatively the observed size correlation. We discuss below what can potentially account for the remaining size correlation.

Since \emph{Kepler} can only detect transiting planets above certain S/N threshold, it is very likely that many of the \emph{Kepler} multi-planet systems may contain additional undetectable planets. Outside the period limit that \emph{Kepler} can probe ($\sim1\,$yr), studies have shown that cold giant planets preferentially co-exist with inner small planets \citep{ZhuWu:2018,Bryan:2019,Herman:2019}, the inclusion of which will certain break the size similarity pattern.

Inside the \emph{Kepler} period domain, there are also signs for additional planets. Dynamical studies have suggested that majority of the \emph{Kepler} multi-planet systems are not fully packed if the detected planets are all the planets in the system. According to \citet{FangMargot:2013} $\sim$55\% of systems with at least four detected planets can contain additional intervening planets without leading to dynamical instability, and the fraction is even higher for systems with two or three detected planets (see also \citealt{PuWu:2015}). Furthermore, the fact that \emph{Kepler} planet detections pile up toward the detection threshold
\footnote{The detection efficiency of \emph{Kepler} pipeline depends on the transit S/N. As \citet{Ciardi:2013} have shown with some earlier Kepler sample (see their figure 1), Kepler detection is only complete for SNR$\gtrsim$25. Therefore, when the incompleteness of the detection pipeline is taken into account, S/N of $\sim$20 is still at the edge of the detection threshold.}
is also suggesting that smaller and undetectable planets do exist.

An intervening undetectable planet between two detectable ones likely also transits and has a smaller (compared to the detection threshold) size. What the addition of such a smaller intervening planet does to the radius correlation (e.g., left panel of Figure~\ref{fig:radius}) is two-fold. First, one planet pair that shows strong correlation is removed and then, two planet pairs that show much weaker correlation are added. The combined consequence is that the size correlation is reduced significantly. To demonstrate this effect, one would like to increase \emph{Kepler}'s sensitivity to recover the smaller planets. This is obviously not practical, so I turn to the opposite direction.
\footnote{This test was originally suggested by Xi Zhang.}
I increase the S/N threshold used in the statistical test, which is equivalent as increasing stellar noises and thus lowering \emph{Kepler}'s sensitivity, and then measure the size correlation in the same way. As shown in Figure~\ref{fig:correlations}, the size correlation becomes stronger in such ``down-graded'' \emph{Kepler} missions. This is aligned with our speculation and suggests that, in a superior \emph{Kepler} mission which can detect much smaller planets and thus is less affected by detection biases, the size correlation should be much weaker.

Another way to show the influence of additional planets on the size correlation is to restrict to high-multiple systems that are less likely to contain additional planets because of stability requirement. I only include systems with at least four CKS planets and repeat the same statistical tests. The results are shown in Figure~\ref{fig:compact}. This time the distribution of the correlation coefficients from statistical tests is statistically closer to the observed value. This again confirms that the missing planets do have an effect on the size correlation.

\section{On the period ratio uniformity} \label{sec:pratio}

\begin{figure*}
\epsscale{1.1}
\plotone{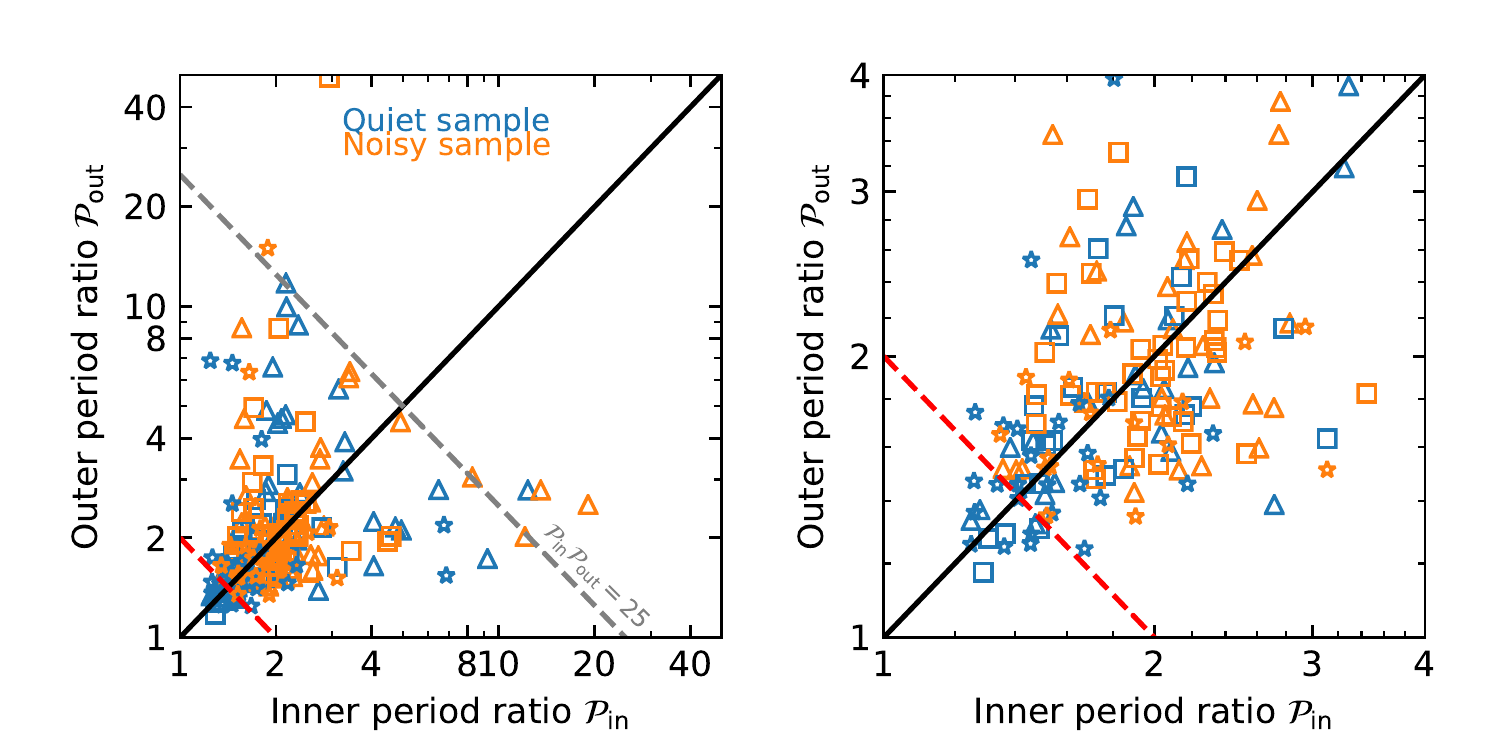}
\caption{The outer period ratio versus the inner period ratio for all triplets (left panel) and the subset with both period ratios $<4$ (right panel). We use triangle, square and asterisk for planet triplet from 3-tranet, 4-tranet and $\ge5$-tranet systems. \citet{Weiss:2018} only used planet triplets with both period ratios $<4$ (i.e., those in the right panel). However, such a square cut is not physically motivated, since the detection probability of a triplet scales as $\mathcal{P}_{\rm in} \mathcal{P}_{\rm out}$. The gray dashed line in the left panel shows an example. We also label with different colors the planet triplets from two stellar samples to highlight their different distributions. In particular, there is no planet triplet with $\mathcal{P}_{\rm in} \mathcal{P}_{\rm out}<2$ (i.e., to the lower left of the red dashed lines).
\label{fig:pratio}}
\end{figure*}

\begin{figure*}
\epsscale{1.1}
\plotone{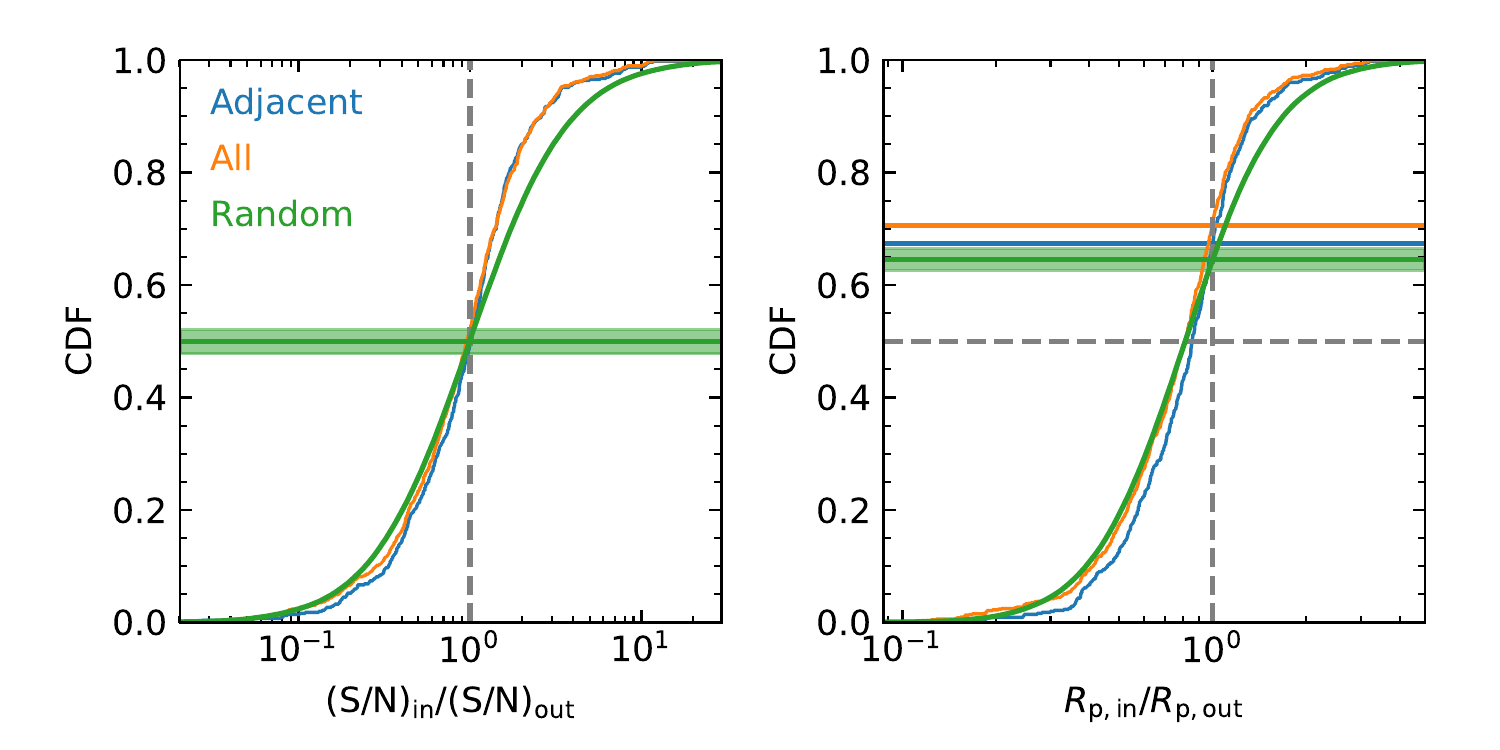}
\caption{Cumulative distributions of the transit S/N ratio (left panel) and the radius ratio (right panel) between planets in a pair. The ratio is specified as the property of the inner to the property of the outer. The two ways of forming planet pairs are studied. The blue curves use only pairs from adjacent planets, whereas the orange curves include pairs from non-adjacent planets. The green curves are results from randomly sampling the S/N distribution. The vertical dashed lines mark the boundary where the quantity of the inner equals the quantity of the outer, and the horizontal solid lines with colors mark the values at which the curves meet these equalities. The green regions mark the $1\sigma$ confidence interval, derived from 1000 realizations of the random sampling, and the horizontal dashed lines mark the median. On average the inner tranet has the same S/N as the outer one, which is what one expects from the random sampling. This naturally leads to an excess of pairs with $R_{\rm p,in}<R_{\rm p, out}$ (i.e., the size-location correlation).
\label{fig:size_hierarchy}}
\end{figure*}

\citetalias{Weiss:2018} also claimed that the spacings between planets, measured by the period ratios, are correlated in systems with at least three CKS planets. To reach this conclusion, they first identified all CKS planet triplets, which consisted of all the CKS planets in three-planet systems and three consecutive planets in higher-multiple systems. For each planet triplet they computed the period ratio between the inner two planets, $\mathcal{P}_{\rm in}$, and the period ratio between the outer two planets, $\mathcal{P}_{\rm out}$. Considering the incomplete sensitivity to large period ratios, \citetalias{Weiss:2018} only included the planet triplets whose $\mathcal{P}_{\rm in}<4$ and $\mathcal{P}_{\rm out}<4$. Then they computed the Pearson $r$ coefficient between the two variables $\log{\mathcal{P}_{\rm in}}$ and $\log{\mathcal{P}_{\rm out}}$ and found $r=0.46$. To assess the significance of this correlation, they generated synthetic systems, in which the period ratios were randomly drawn from the overall period ratio distribution, and found that the correlations in these simulated samples were systematically much smaller than the observed one.

This approach may produce biased results in two ways. First, the cut at period ratio $\mathcal{P}=4$ is fairly arbitrary and not physically motivated. In the left panel of Figure~\ref{fig:pratio} I show the $\mathcal{P}_{\rm out}$ versus $\mathcal{P}_{\rm in}$ for all planet triplets in the \citetalias{Weiss:2018} sample. One can see that the detection limit at large period ratios is diagonal rather than flat. This can be well explained by the detectability of these multi-planet systems. If the planets in the same system have coplanar or nearly coplanar orbits, the detectability of all planets only depends on the orbital period of the outermost planet, $P_{\rm outermost}$. For a planet triplet, this detection threshold  scales as $\propto \mathcal{P}_{\rm in}\mathcal{P}_{\rm out}$. In the left panel of Figure~\ref{fig:pratio}, I plot the line that corresponds to $\mathcal{P}_{\rm in}\mathcal{P}_{\rm out}=25$, and it roughly agrees with the upper boundary of all data points. Adopting this physically motivated detection threshold, I find the Pearson $r=0.21$ between $\log{\mathcal{P}_{\rm in}}$ and $\log{\mathcal{P}_{\rm out}}$. Restricting to $\mathcal{P}_{\rm in}\mathcal{P}_{\rm out}<16$ gives $r=0.15$. Both correlations are much weaker than that given by \citetalias{Weiss:2018}.

The second issue is the varying detection threshold of transit. At first glance, the S/N, as given in Equation~(\ref{eqn:snr}), has only weak dependence on orbital period and no explicit dependence on the period ratio. The period ratio comes into play via the dynamical stability requirement. The stability boundary is typically measured in the number ($K$) of mutual Hill radii, $r_{\rm H}$,
\begin{equation}
a_2 - a_1 = K \cdot r_{\rm H},\quad r_{\rm H} \equiv \frac{a_1+a_2}{2} \left(\frac{m_1+m_2}{3}\right)^{1/3}\ ,
\end{equation}
where $a_i$ and $m_i$ are the semi-major axis and mass of the inner ($i=1$) and outer ($i=2$) planets, respectively. For simplicity, we further assume $m_1\approx m_2 \approx M_\oplus (R_{\rm p}/R_\oplus)^3$. Note that this is not valid in general, but it is acceptable for the planet pairs that are just above the detection threshold and close to the instability limit. Then with Kepler's third law we can have a rough scaling between the planetary size and the critical period ratio for dynamical stability
\begin{equation}
\mathcal{P} = \frac{P_2}{P_1} \approx 1+0.019K \left(\frac{\rp}{R_\oplus}\right)\ .
\end{equation}
Below we adopt $K=20$, although in reality the threshold on $K$ also depends on many factors, such as individual planet masses, eccentricities and mutual inclinations, etc (e.g., \citealt{Chambers:1996,Zhou:2007}; see \citealt{PuWu:2015} for a detailed discussion). As Figure~\ref{fig:sample} shows, the smallest planet detectable around a typical ``noisy'' star is $1R_\oplus$, for which the stability threshold is $\mathcal{P}_{\rm crit,1}\approx1.4$. The smallest detectable around a typical ``quiet'' star, in contrast, is $0.5R_\oplus$, with a stability threshold of $\mathcal{P}_{\rm crit,2}\approx1.2$. This varying threshold is visible in the lower middle panel of Figure~\ref{fig:sample} as well as the right panel of Figure~\ref{fig:pratio}. Again for demonstration purposes I have differentiated the planet triplets from the quiet sample and noisy sample with different colors. No planet triplets from the noisy sample are below the red dashed line, which denotes $\mathcal{P}_{\rm in} \mathcal{P}_{\rm out}=2\approx \mathcal{P}_{\rm crit,1}^2$, whereas planet triplets from the quiet sample can extend further down to $\mathcal{P}_{\rm crit,2}^2$. This varying stability threshold was not taken into account in \citetalias{Weiss:2018}.

The varying stability threshold applying to an inhomogeneous stellar sample naturally leads to a spacing correlation, as is illustrated in the right panel of Figure~\ref{fig:schematic}. Note the similarity between this plot and the right panel of Figure~\ref{fig:pratio}. Generating synthetic planetary systems that meet all detection thresholds of individual planets and of the triplet as well as the stability threshold is not trivial, so I cannot assess quantitatively the impact of this effect on the Pearson $r$ coefficient. However, as a qualitative check, if only planet triplets from the noisy sample are used, I have $r=0.25$ even with the square cut at $\mathcal{P}=4$. This is much smaller than what one has ($r=0.46$) if both noisy and quiet samples are used.

\section{On the size ordering} \label{sec:size_hierarchy}

\citet{Lissauer:2011} and \citet{Ciardi:2013} first noticed that the \emph{Kepler} multi-planet systems show a size-location correlation. Specifically, the larger planet in any planet pair is most often the one with the longer period. To check the statistical significance against observation biases, these authors compared the radius ratio distributions between observation and simulation. In generating simulated planet pairs, they randomly drew radii from the observed radius distribution and then, to mimic their selection procedure, eliminated those which would not be detected if either of the planets at the orbital period of the other one fell below the specified S/N threshold. Their simulated radius ratio distribution showed equal number of planet pairs with $R_{\rm p,in}<R_{\rm p,out}$ and $R_{\rm p,in}>R_{\rm p,out}$. \citet{Ciardi:2013} also performed several other tests, including using different S/N thresholds and maximum periods. The size ordering was always observed. Therefore, it was concluded that the size-location correlation has a physical origin.

This statistical approach suffers the same issue as the \citetalias{Weiss:2018} one. Using the CKS multi-planet sample, I show in Figure~\ref{fig:size_hierarchy} the cumulative distributions of transit S/N ratios and radius ratios between planets in pairs. Similar to what \citet{Lissauer:2011} and \citet{Ciardi:2013} found, here I also have more than 60\% of planet pairs showing the so-called size-location correlation: $R_{\rm p,in}<R_{\rm p,out}$. However, the S/N ratio is on average unity, suggesting that the transit signal of the inner planet is as strong as that of the outer one. This is what one expects if randomly pairs up the transit S/N values from the observed S/N distribution. See the green curve in the left panel of Figure~\ref{fig:size_hierarchy}. Given the relation between $\snr$ and radius (Equation~(\ref{eqn:snr})), one can derive the radius ratio from the $\snr$ ratio
\begin{equation}
\frac{(\snr)_{\rm in}}{(\snr)_{\rm out}} \approx \left(\frac{R_{\rm p,in}}{R_{\rm p,out}}\right)^2 \left(\frac{P_{\rm in}}{P_{\rm out}}\right)^{-1/3} .
\end{equation}
In the above approximation we ignore the contribution from impact parameters. Because of the term involving the period ratio, two transit signals with equal S/N naturally lead to a pair of planets with $R_{\rm p,in}<R_{\rm p,out}$, that is, the size-location correlation.

The distributions from randomly sampling the S/N distribution do not match the observed distributions perfectly, in particular in the range $1<(\snr)_{\rm in}/(\snr)_{\rm out}<4$, or equivalently $1<R_{\rm p,in}/R_{\rm p,out}<2$. There are two possible reasons. First, transiting planets in some pairs do show a weak size correlation. This is also suggested in Section~\ref{sec:radius}, as the random $\snr$ sampling cannot fully explain the observed correlation strength (Figure~\ref{fig:histograms}). However, as is also suggested in Section~\ref{sec:radius}, this can possibly be explained by transit bias, as we are not detecting all the planets in the same system. Regardless, the fraction of planet pairs that show the size correlation is likely a small fraction ($\lesssim5\%$). Otherwise it would require a very fine-tuned period ratio distribution to push the median $\snr$ ratio to almost exactly unity.

Another possible reason that can account for the deviation is some subtle detection bias in the planet search pipeline. When S/N values are randomly paired up, it is implicitly assumed that the detection efficiency does not depend on parameters other than $\snr$. This is not entirely true in reality. For the same value of S/N, the detection efficiency decreases gradually with the orbital period (see, e.g., Figure~9 of \citealt{Thompson:2018}). This effect biases against planet pairs with large $R_{\rm p,in}/R_{\rm p,out}$, the type of pairs that are in short for the observational distribution to match the simulated one. Future detailed studies are needed to quantify this effect.

In short, the observed size-location correlation in \emph{Kepler} multi-planet systems can be mostly, if not fully, explained by detection biases. It is possible that some planet pairs do have ordered sizes, but they only consist of a small fraction ($\lesssim 5\%$) of all planet pairs.

\section{Discussion} \label{sec:discussion}

\begin{figure}
\epsscale{1.2}
\plotone{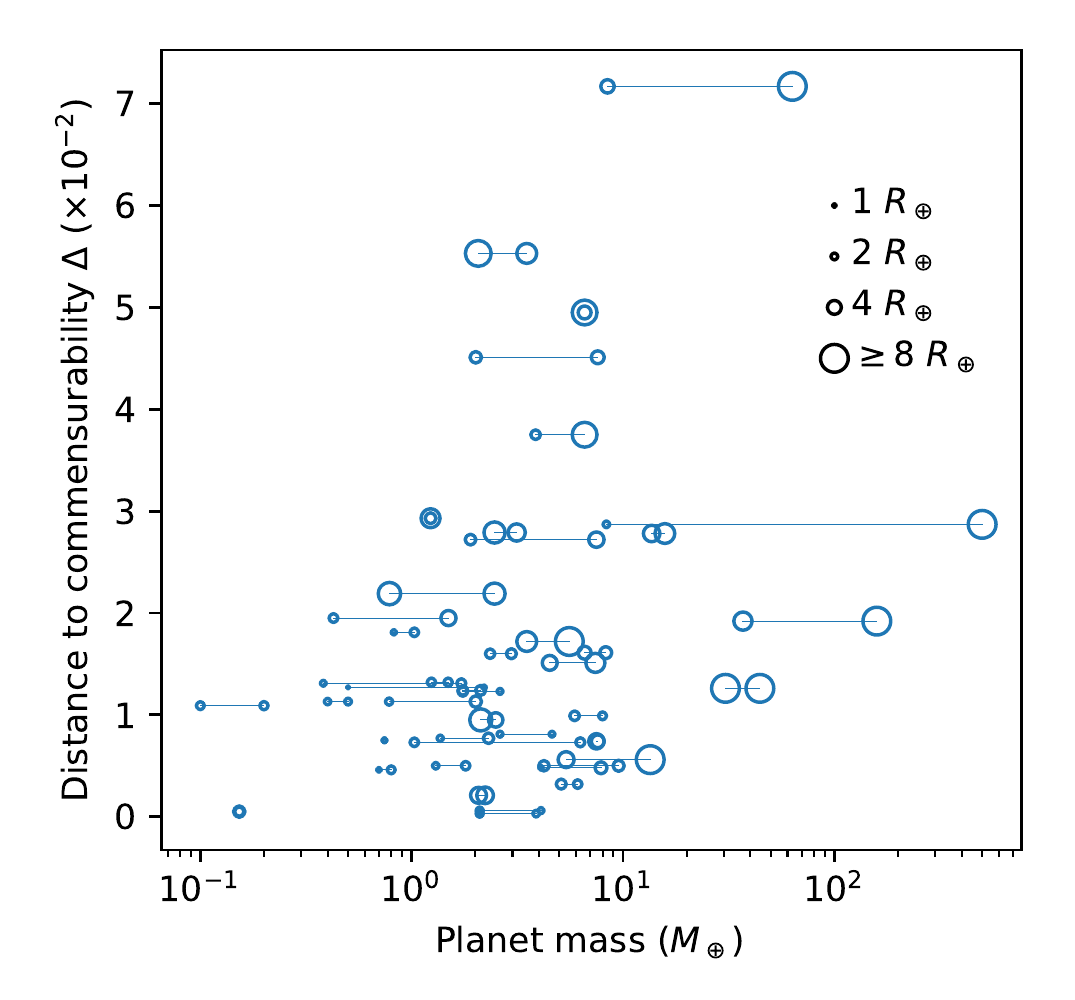}
\caption{An illustration of the planet pairs used in \citet{Millholland:2017} for claiming the intra-system mass uniformity. For each pair, I show on the $x$-axis the masses of individual planets and on the $y$-axis the distance to exact period commensurability, $\Delta$, given by Table~2 of \citet{Hadden:2017}. The size of the symbol reflects the radius of the planet. As $\Delta$ decreases, lower masses can be detected via the TTV technique. This detection bias is not taken into account by \citet{Millholland:2017} in constructing synthetic systems.
\label{fig:mass_bias}}
\end{figure}

In this work, I re-examine several claims about the relative sizes and spacings between \emph{Kepler} planets around the same host. I make use of the observed transit S/N values, because they are observationally more fundamental than other parameters such as the planet radius. I present several findings:
\begin{itemize}
\item The apparently similar sizes of planets in the same \emph{Kepler} system can be largely explained by the projection of the same S/N cut onto different stellar properties.
\item The apparently correlated spacings, measured in period ratios, between adjacent planet pairs in systems with at least three detected planets are partially due to the arbitrary upper limit that \citetalias{Weiss:2018} imposed on the period ratio and partially due to the varying stability threshold in different stellar samples.
\item The observed size-location correlation can be explained by the projection of the same S/N onto different values of orbital period. As far as the transit detection is concerned, the inner and the outer transiting planets on average have similar S/N values.
\end{itemize}

The claim of intra-system mass uniformity by \citet{Millholland:2017} suffers very similar issues. Below I draw the analogy between \citet{Millholland:2017} and \citetalias{Weiss:2018} studies and defer a more quantitative analysis for future works. The analysis of \citet{Millholland:2017} was performed on a sample of 89 planets from 37 Kepler systems, whose masses were constrained by \citet{Hadden:2017} through TTV.
\footnote{\citet{Millholland:2017} had 8 planets with only mass upper limits, as indicated in \citet{Hadden:2017}, in their sample: Kepler-23 d, Kepler-24 e, KOI-115.03, Kepler-105 b, Kepler-114 b, Kepler-114 d, Kepler-310 c, and KOI-427.01. It is not appropriate to treat mass upper limits and mass measurements in the same way. The removal of these planets reduces the number of transiting planets in four systems (Kepler-105, 114, 310, 549) down to one, so these systems are excluded from the sample. In the end I have 77 planets from 33 systems.}
Whether or not a TTV mass measurement can be made is more directly related to the TTV amplitude than to the planet mass. For a pair of planets, the TTV amplitude is generally dependent on the distance from the period commensurability, $\Delta$, which is defined as \citep{Lithwick:2012}
\begin{equation}
\Delta \equiv \left| \frac{P_{\rm out}}{P_{\rm in}} \frac{J^\prime}{J} -1 \right|.
\end{equation}
Here $P_{\rm in}$ and $P_{\rm out}$ are the orbital periods of the inner and outer planet in a TTV pair, respectively, and $J^\prime/J$ is the closest small integer ratio for $P_{\rm in}/P_{\rm out}$. Other things being equal, a smaller $\Delta$ means that a lower planet mass can be measured from TTV. This makes the $\Delta-m_{\rm p}$ relation (Figure~\ref{fig:mass_bias}) somewhat analogous to the $\cdpp-\rp$ relation (lower left panel of Figure~\ref{fig:sample}). Consequently, \citet{Millholland:2017} reshuffling the planet mass is similar to \citetalias{Weiss:2018} bootstrapping the planet radius. As shown in Section~\ref{sec:radius}, this approach leads to biased results.

Therefore, the so-called intra-system uniformity and the size ordering effect that appear in the \emph{Kepler} multi-planet systems can be mostly, if not entirely, explained by observational biases. As far as the data is able to inform, the physical properties of one \emph{Kepler} planet are largely independent of the properties of both its siblings and the parent star.

So far the analysis has been done on the \emph{Kepler} multi-planet systems, but the same conclusion likely applies to all \emph{Kepler} planets. It is true that over half of the transiting planets were found in systems with only one transiting planets (i.e., single-tranet systems). However, this is most likely a result of selection effect, as \emph{Kepler} only detects planets that transit the host star. The orbital properties, such as eccentricity and mutual inclination, of single-tranets and multi-tranets are different \citep[e.g.,][]{Xie:2016,Zhu:2018,VanEylen:2019}, but this does not necessarily mean that their physical properties are different as well. In fact, studies have shown that the planetary properties and the stellar properties of single-tranet and multi-tranet systems are similar \citep[e.g.,][]{MunozRomero:2018,Zhu:2018,Weiss:2018b}, suggesting they are likely the same population. Nevertheless, future ground-based radial velocity observations will be able to tell whether or not this is true.

A recent work by \citet{He:2019} applied the full forward modeling method to study the \emph{Kepler} multi-planet systems. The authors generated multi-planet systems following specific prescriptions, passed the simulated systems through simplified \emph{Kepler} detection pipeline, and compared their simulated planet catalogs to the real catalog. The authors find a better match in a combination of selected observables between the simulated and real catalogs once the periods and radii of planets around the same host are assumed to be correlated. However, it is unclear whether the improvement in the fit is due to physical correlations or some artifacts in the model. In fact, as their best-fit models show (Figures~3-5 of \citealt{He:2019}), the match to transit depth and transit depth ratio distributions, the most relevant ones for the size correlation, is not improved once the size correlation is introduced.
\footnote{Their best-fit ``clustered periods and sizes'' model actually gives worse match to the observed distributions of transit depth and transit depth ratios than their best-fit ``non-clustered'' model does.}
This is in agreement with the conclusion of the current paper that there is no evidence for the size correlation. The spacing correlation is a much more complicated issue in such a full forward modeling approach. To give a specific example, how to generate stable multi-planet systems remains an unsolved problem. The critical spacing for long-term stability depends on many factors, including the number of planets (i.e., multiplicity, \citealt{Funk:2010}) and orbital properties (i.e., eccentricity and mutual inclination, \citealt{PuWu:2015}), the latter of which have also been shown to be multiplicity-dependent \citep{Xie:2016,Zhu:2018}. Using a fixed $K$ value for all systems, as was done in \citet{He:2019}, is not realistic. More work is needed.

The conclusion that the properties of \emph{Kepler} planets are largely independent of the properties of their siblings and the parent star has theoretical implications. Either the formation of \emph{Kepler} planets had almost no requirement for their birth environment, or the (likely) chaotic evolution erased their memory of the initial condition. This latter scenario is more likely once several other pieces of evidence are put together. \emph{Kepler} multi-planet systems are shown to be dynamically compact \citep{PuWu:2015} and have very diverse compositions \citep{Wu:2013,Marcy:2014,Hadden:2017}. The nearly flat period ratio distribution, arisen from either \textit{in situ} formation \citep{Petrovich:2013,Wu:2019} or breaking the chain of resonances after migration \citep{Izidoro:2017}, also points to a stage of dynamical instability. Finally, \emph{Kepler} planets are shown to be strongly correlated with outer giant planets \citep{ZhuWu:2018,Bryan:2019,Herman:2019}. The planet-planet scatterings that are responsible for the large eccentricities of the cold giant planets \citep{Chatterjee:2008,Juric:2008} can easily drive dynamical instabilities in the inner system.
 
As the orbital velocity far exceeds the escape velocity for the majority of \emph{Kepler} planets, the encounters between planets during the dynamical evolution can significantly revise their physical properties, thus removing the imprint of the initial formation conditions. In other words, it will be difficult to infer the initial conditions from the properties of current \emph{Kepler} planets. A different conclusion was reached in \citet{Kipping:2018} from studying the size orderings of \emph{Kepler} planets. \citet{Kipping:2018} assumed that the observed orderings are physical and free from observational biases. This is not true, as the present study has shown.

\acknowledgements
I would like to thank Subo Dong, Cristobal Petrovich, Yanqin Wu, Norm Murray, and Eve Lee for discussions, and particularly Xi Zhang for the suggestion about the new test in the size correlation section. I also thank the anonymous referees for comments and suggestions on the manuscript. I also thank Lauren Weiss for kindly sharing her code which helps reproduce their results.
W.Z. was supported by the Beatrice and Vincent Tremaine Fellowship at CITA.


\end{CJK*}
\end{document}